\documentclass{iopconfser}

\usepackage[utf8]{inputenc}
\usepackage[T1]{fontenc}

\usepackage{amsmath}

\usepackage{amsfonts}
\usepackage{mathrsfs}

\usepackage{amssymb}
\usepackage{units}

\usepackage{graphicx}

\usepackage{placeins}

\usepackage{verbatim}

\usepackage{accents}

\usepackage{multirow}

\usepackage{color}

\usepackage{orcidlink}

\newcommand{\mrm}{\mathrm}

\makeatletter

\let\old@dmathbeg\[
\let\old@dmathend\]

\newcommand{\rovnec}[1]{\old@dmathbeg#1\old@dmathend}
\newcommand{\rovcis}[2]{\begin{equation}#1\label{#2}\end{equation}}
\newcommand{\drovcis}[2]{\begin{equation}\begin{split}#1\end{split}\label{#2}\end{equation}}
\newcommand{\drovnec}[1]{\begin{equation*}\begin{split}#1\end{split}\end{equation*}} 
\newcommand{\provcis}[1]{\begin{align}#1\end{align}}
\newcommand{\provnec}[1]{\begin{align*}#1\end{align*}}

\newcommand{\rov}{\@ifstar\rovnec\rovcis}
\newcommand{\drov}{\@ifstar\drovnec\drovcis}
\newcommand{\prov}{\@ifstar\provnec\provcis}

\newcommand{\vast}{\bBigg@{4}}
\newcommand{\Vast}{\bBigg@{5}}

\makeatother

\DeclareMathOperator{\diffbold}{\mathbf{d}}
\newcommand{\bd}{\diffbold\!}

\newcommand{\mbs}{\boldsymbol}

\newcommand{\iPi}{{\mit\Pi}}

\DeclareFontEncoding{LGR}{}{}

\DeclareMathAlphabet{\msi}{OT1}{cmss}{m}{it}

\DeclareMathAlphabet{\mgr}{LGR}{cmr}{m}{n}

\newcommand{\rpi}{\mgr{p}}

\renewcommand{\[}{\left[}
\renewcommand{\]}{\right]}
\renewcommand{\l}{\left}

\newcommand{\ri}{\right}

\newcommand{\zrov}{{}\\{}}

\newcommand{\res}[2]{\l.#1\ri|_{#2}}
\newcommand{\lbl}{\label}

\newcommand{\qt}[1]{``#1''}

\newcommand{\pmu}[1]{p^\mu_{(#1)}}
\newcommand{\pt}[1]{p_t^{(#1)}}
\newcommand{\pf}[1]{p_{\vphantom{t}\varphi}^{(#1)}}
\newcommand{\adet}[1]{\begin{vmatrix}#1\end{vmatrix}}

\newcommand{\PRD}[2]{\href{https://dx.doi.org/10.1103/PhysRevD.#1.#2}{{\it Phys. Rev.} D {\bf #1} #2}}
\newcommand{\PRL}[2]{\href{https://dx.doi.org/10.1103/PhysRevLett.#1.#2}{{\it Phys. Rev. Lett.} {\bf #1} #2}}
\newcommand{\arXiv}[1]{\href{https://arxiv.org/abs/#1}{arXiv:#1}}

\renewcommand{\(}{\left(}
\renewcommand{\)}{\right)}

\begin{document}

\title{Energy extraction from electrovacuum black holes via production of pairs of oppositely charged particles
}

\author{F Hejda\,\orcidlink{0000-0002-4157-6730}$^{1, 2}$}

\affil{$^1$Centro de Astrofísica e Gravitação -- CENTRA, Departamento de Física, Instituto Superior Técnico -- IST, Universidade de Lisboa -- UL, Avenida Rovisco Pais 1, 1049-001 Lisboa, Portugal}
\medskip
\affil{$^2$CEICO, Institute of Physics of the Czech Academy of Sciences, Na Slovance 1999/2, 182 21 Prague 8, Czech Republic}

\email{filip.hejda@tecnico.ulisboa.pt}

\begin{abstract}
We consider collisional Penrose process for charged, rotating black holes together with a simple model of pair creation, in which two oppositely charged particles are produced in a collision of two neutral particles. We highlight that significant energy extraction is possible without assuming fine-tuning or extremality as long as the escaping particles are sufficiently charged.
\end{abstract}

\section{Introduction}

\enlargethispage{\baselineskip}

Astrophysical viability of the Penrose process has been questioned, because in order to achieve energy extraction, the fragments of the disintegrating particle must have a relative velocity of more than half the speed of light, which is very restrictive \cite{Wald74a, BarPrTeu}. However, the restriction can be circumvented through inclusion of electromagnetic interaction; this has been shown for a disintegration of a neutral particle into oppositely charged fragments near a weakly magnetised black hole \cite{WaDhuDa}. Another possible generalisation is to consider particle collisions instead, in which case the high relative velocity of the final particles might arise as a natural result of a high-energy collision \cite{PirShK, PirSh}. Much of the detailed research into such collisional Penrose process is focused on situations in which 
a collision involving a fine-tuned particle close to an extremal black hole gives rise to arbitrarily large centre-of-mass collision energy, i.e.\ generalised BSW effect \cite{BSW, Zasl10, Zasl11a, a2}. Yet even in such idealised cases, it turns out that the energy that can be extracted is subject to strict upper bounds \cite{BPAH, HaNeMi, Schnitt14}, which are however lifted whenever both the black hole and the escaping particle are charged \cite{Zasl12c, a3, a4}. Therefore, electromagnetic interaction is the true enabling factor, like in \cite{WaDhuDa}.

\section{The algebraic solution to the conservation of momentum at the instant of collision}

Let us start by considering a general stationary and axially symmetric electrovacuum spacetime with metric $\mbs g$ and electromagnetic potential $\mbs A$ given by
\prov{\mbs g&=-N^2\bd t^2+g_{\varphi\varphi}\(\bd\varphi-\omega\bd t\)^2+g_{rr}\bd r^2+g_{\vartheta\vartheta}\bd\vartheta^2&\mbs A&=A_t\bd t+A_\varphi\bd\varphi}
The black-hole horizon located at $r=r_\mrm{H}$ corresponds to $N=0$. 
Provided that $\mbs g$ and $\mbs A$ 
also obey a reflection symmetry with respect to the equatorial \qt{plane} $\vartheta=\nicefrac{\rpi}{2}$, 
charged test particles with initial conditions $\vartheta=\nicefrac{\rpi}{2}$, $p^\vartheta=0$ will stay within this hypersurface. 
Motion of such an equatorial particle with mass $m$, energy $E$, angular momentum $L$ and charge $q$ is governed by the following equations:
\prov{p^t&=\frac{X}{N^2}&p^\varphi&=\frac{\omega X}{N^2}+\frac{p_\varphi}{g_{\varphi\varphi}}&p^r={\frac{\sigma P}{N\sqrt{g_{rr}}}}\lbl{eomp}}
Here $\sigma=\pm1$ determines the direction of the radial motion and functions $X$, $P$, $p_t$ and $p_\varphi$ correspond to
\prov{X&=-p_t-\omega p_\varphi&P&=\sqrt{X^2-N^2\(m^2+\frac{p_\varphi^2}{g_{\varphi\varphi}}\)}&p_t&=-E-qA_t&p_\varphi&=L-qA_\varphi\lbl{Xdef}}

The motion is allowed whenever $P^2\geqslant0$. Assuming $g_{\varphi\varphi}>0$, this implies that $X>0$ and $X<0$ are disjoint domains and $X$ cannot change sign during motion of a single particle. In order to preserve causality, we shall enforce motion forward in time and restrict ourselves to particles with $X>0$. Then we can say that so-called \emph{supercritical} particles with $X_\mrm{H}<0$ are forbidden to fall into the black hole, because they must hit $P=0$, which corresponds to a turning point, before they could reach the horizon.

Let us now consider a scattering process in which initial particles $1$ and $2$ collide and produce final particles $3$ and $4$. We shall impose conservation of charge $q_1+q_2=q_3+q_4$ and conservation of momentum $\pmu{1}+\pmu{2}=\pmu{3}+\pmu{4}$. Let us define $P_0$, $\sigma_0$, $\pt{0}$ and $\pf{0}$ to parameterise the centre-of-mass frame
\prov{\sigma_0P_0&=\sigma_1P_1+\sigma_2P_2&\pt{0}&=\pt{1}+\pt{2}&\pf{0}&=\pf{1}+\pf{2}}
Then the conservation of momentum can be recast into the following equation:
\drov{\sigma_0P_0=\sigma_3&\sqrt{\(\pt{3}+\omega\pf{3}\)^2-N^2\[m_3^2+\frac{1}{g_{\varphi\varphi}}\(\pf{3}\)^2\]}+\zrov+\sigma_4&\sqrt{\[\pt{0}-\pt{3}+\omega\(\pf{0}-\pf{3}\)\]^2-N^2\[m_4^2+\frac{1}{g_{\varphi\varphi}}\(\pf{0}-\pf{3}\)^2\]}}{mastercons}
Curiously enough, this equation has a relatively simple algebraic solution. This was first indicated in \cite{Wald74a, WaDhuDa} and recently studied in detail in \cite{Zasl19, Zasl23, Zasl24}. Let us further define $m_0$ and $\iPi_3$ by
\prov{N^2m_0^2&=\(\pt{0}+\omega\pf{0}\)^2-P_0^2-\frac{N^2}{g_{\varphi\varphi}}\(\pf{0}\)^2&\iPi_3&=m_0^2+m_3^2-m_4^2}
The algebraic solution to \eqref{mastercons} can be expressed as a generic quadratic implicit equation in two variables
\rov{\msi A\(\pt{3}\)^2+2\msi B\pt{3}\pf{3}+\msi C\(\pf{3}\)^2+2\msi D\pt{3}+2\msi E\pf{3}+\msi F=0}{impsol}
with coefficients given by
\prov{\msi A&=\(\pf{0}\)^2+g_{\varphi\varphi}m_0^2&\msi B&=-\(\pt{0}\pf{0}+g_{t\varphi}m_0^2\)&\msi C&=\(\pt{0}\)^2+g_{tt}m_0^2\\\msi D&=-\frac{\iPi_3}{2}\(g_{\varphi\varphi}\pt{0}-g_{t\varphi}\pf{0}\)&\msi E&=-\frac{\iPi_3}{2}\(g_{tt}\pf{0}-g_{t\varphi}\pt{0}\)&\msi F&=g_{\varphi\varphi}\(P_0^2m_3^2+\frac{N^2}{4}\iPi_3^2\)}
In order to determine the geometric interpretation of \eqref{impsol}, one shall construct the following two determinants out of its coefficients (cf. \cite{Wolfram}):
\prov{\msi J&=\adet{\msi A&\msi B\\\msi B&\msi C}=g_{\varphi\varphi}P_0^2m_0^2&\msi I&=\adet{\msi A&\msi B&\msi D\\\msi B&\msi C&\msi E\\\msi D&\msi E&\msi F}=-\frac{g_{\varphi\varphi}^2P_0^4}{4}\(\iPi_3^2-4m_0^2m_3^2\)}
Since $\msi J>0$, \eqref{impsol} must represent an ellipse. 
However, it will be defined in real numbers only if $\msi I$ has the opposite sign to $\msi A$ and $\msi C$. Therefore, we get a requirement $\iPi_3^2>4m_0^2m_3^2$. Upon restriction to all particles moving forward in time, this can be equivalently restated as $m_3+m_4<m_0$. 
Because $m_0$ equals the centre-of-mass collision energy $E_\mrm{c.m.}$ (cf. \cite{a2}), 
we see that $E_\mrm{c.m.}$ serves as an upper bound for the masses of the produced particles. 

The solution \eqref{impsol} implies the following bounds on the values of $\pt{3}$ and $\pf{3}$ (cf. \cite{Zasl19, Zasl23}):
\prov{\pm\pt{3}&\leqslant\frac{1}{2m_0^2}\(\sqrt{\msi C}\sqrt{\iPi_3^2-4m_0^2m_3^2}\pm\iPi_3\pt{0}\)\lbl{ptmax}\\\pm\pf{3}&\leqslant\frac{1}{2m_0^2}\(\sqrt{\msi A}\sqrt{\iPi_3^2-4m_0^2m_3^2}\pm\iPi_3\pf{0}\)\lbl{pfmax}}

\section{Conclusions for production of oppositely charged pairs in collisions of neutral particles}

In order to extract energy from the black hole, one of the produced particles must escape, 
whereas the other one must fall inside. Calculating the escape conditions can be complicated in general. 
Nevertheless, it has been repeatedly shown that the largest energy extraction is achieved 
for initially incoming particles that get reflected (see \cite{a4} and references therein). 
Here we shall consider a particular model process, for which we can determine in a simple way the possibility to produce particles with $X_\mrm{H}<0$ (which get reflected due to being kinematically forbidden to fall inside, as noted above). In particular, let us study a near-horizon collisional process characterised by the following assumptions: (i)~the initial particles are uncharged (ii) the energies of the initial particles are not much larger than the masses of the final particles (iii) for each of the final particles, the magnitude of the product of its charge with the charge of the black hole is much larger than the product of its mass with the mass of the black hole. (Note that the final particles need to be oppositely charged due to charge conservation and that magnitudes of their charges will be much larger than their masses since the magnitude of the black hole charge cannot be larger than its mass.)
The physical motivation for this setup is to model pair creation via a collision of photons. However, the conclusion below does not require the initial particles to be massless. 

\enlargethispage{\baselineskip}

From the definition of $X$ \eqref{Xdef}, we can express the value of $X_3$ at the horizon using its value at the point of inception (marked by $\mrm{C}$) minus two difference terms
\rov{X_3^\mrm{H}=X_3^\mrm{C}-\Delta\omega\(\pf{3}\)_\mrm{C}-q_3\xi}{X3Hdiff}
Here $\Delta\omega=\omega_\mrm{H}-\omega_\mrm{C}$, whereas $\xi$ is given by
\rov{\xi=A_t^\mrm{C}-A_t^\mrm{H}+\omega_\mrm{H}\(A_\varphi^\mrm{C}-A_\varphi^\mrm{H}\)}{}
Let us now assess the balance of the three terms on the r.h.s.\ of \eqref{X3Hdiff} using our assumptions (i)-(iii).
First, uncharged particles that can get close to the horizon must have a specific relation between energy and angular momentum. (In order to reach the horizon from the infinity, the angular momentum of a particle must lie between the values associated with the prograde and retrograde unstable orbits corresponding to the particle's  energy, cf. \cite{BarPrTeu}.) Consequently, values $X_1^\mrm{C}$ and $X_2^\mrm{C}$ will be bounded by some function of energies $E_1$ and $E_2$, and thus not much larger than $m_3$ due to assumption (ii). Recalling that $X$ is just a rescaled component of the momentum \eqref{eomp}, we can see that it is additive and conserved at the point of collision, i.e.\ $X_1^\mrm{C}+X_2^\mrm{C}=X_3^\mrm{C}+X_4^\mrm{C}$. 
Moreover, since $X$ must be positive, we can conclude that $X_3^\mrm{C}<X_1^\mrm{C}+X_2^\mrm{C}$. Hence, $X_3^\mrm{C}$ cannot be much larger than $m_3$. (Similarly, one can infer that neither $\pt{0}$ nor $\pf{0}$ can be much larger than $m_3$, which limits their effect on \eqref{ptmax} and \eqref{pfmax}.)

Turning to the second term on the r.h.s.\ of \eqref{X3Hdiff}, we shall note that for generalised BSW effect near the horizon of an extremal black hole (see, e.g.\ \cite{a4}), $m_0^2\equiv E_\mrm{c.m.}^2$ can grow like $E_\mrm{c.m.}^2\sim\(r_\mrm{C}-r_\mrm{H}\)^{-1}$, and thus \eqref{pfmax} implies that $\(\pf{3}\)_\mrm{C}$ can grow like $\(\pf{3}\)_\mrm{C}\sim\(r_\mrm{C}-r_\mrm{H}\)^{-\frac{1}{2}}$. However, even this fine-tuned growth will be suppressed by the $\Delta\omega$ factor, which must decrease like $\Delta\omega\sim\(r_\mrm{C}-r_\mrm{H}\)$. Therefore, the second term on the r.h.s.\ of \eqref{X3Hdiff} will necessarily be small in the vicinity of the horizon. Consequently, without the $q_3\xi$ term, $X_3^\mrm{H}$ can be made negative only if $X_3^\mrm{C}$ is small. This is why many previous works (e.g.\ \cite{HaNeMi, Zasl12c, a3, a4}) solely focused on processes that lead to production of particles with small $X_3^\mrm{C}$.

Conversely, if we do not assume extremality or fine-tuning and $X_3^\mrm{C}$ will be neither small nor much larger than $m_3$, the second term on the r.h.s.\ of \eqref{X3Hdiff} will be negligible and we shall focus on the $q_3\xi$ term. For a Kerr-Newman black hole with mass $M$, angular momentum $aM$ and charge $Q$, $\xi$ can be evaluated as follows:
\prov{
\res{\mbs A}{\vartheta=\frac{\rpi}{2}}&=-\frac{Q}{r}\(\bd t-a\bd\varphi\)&\xi&=\frac{Qr_\mrm{H}}{2Mr_\mrm{H}-Q^2}\(1-\frac{r_\mrm{H}}{r_\mrm{C}}\)\sim\frac{Q}{M}\(1-\frac{r_\mrm{H}}{r_\mrm{C}}\)\lbl{KNAxi}}
As we can see, 
 assumption (iii) guarantees that all particles produced with the same sign of charge as the black hole will escape, as long as the process happens at a radius larger than some threshold radius $r_\mrm{I}$, which can be   approximated by
\rov{\frac{r_\mrm{I}}{r_\mrm{H}}-1\sim \frac{m_3M}{q_3Q}}{ringest}
Hence, energies of escaping particles produced at $r_\mrm{C}>r_\mrm{I}$ will span the whole range implied by \eqref{ptmax} with \eqref{Xdef}. However, using \eqref{KNAxi}, we can show that the terms on the r.h.s.\ of \eqref{ptmax} are negligible compared to $q_3A_t$ per assumptions (ii) and (iii). We can thus estimate $E_3$ and  the efficiency $\eta$ of the Penrose process 
\prov{E_3&\approx\frac{q_3Q}{r_\mrm{C}}\sim\frac{q_3Q}{M}&\eta&\equiv\frac{E_3}{E_1+E_2}\sim\frac{q_3Q}{m_3M}\lbl{E3eta}}
In so doing, we can also observe that assumption (iii) implies $E_3\gg m_3$ and assumption (ii) further implies $\eta\gg1$.

Now, let us consider an electron, which has $q_3\approx-2\cdot 10^{21}m_3$, and a slightly negatively charged black hole with $Q=-10^{-11}M$, for example. Then all the electrons produced in realisations of our model process will escape, unless the process occurs at a radius smaller than $r_\mrm{I}$, which would be \emph{absurdly} close to the horizon, since \eqref{ringest} implies $\(r_\mrm{I}-r_\mrm{H}\)\sim 10^{-10}r_\mrm{H}$. Using \eqref{E3eta}, we can conclude that in this example the escaping electrons will have energies $E_3\sim 10^{10} m_3$ and the Penrose process will reach efficiency $\eta\sim 10^{10}$.

\section*{Acknowledgments}

The author would like to honour the memory of Jiří Bičák, who opened up this research avenue for him. The author also wants to 
give thanks to Oleg B. Zaslavskii and José P. S. Lemos for collaboration on previous projects studying energy extraction and to Vitor Cardoso and David Hilditch for stimulating discussions.
This work was supported by the European Union and the Czech Ministry of Education, Youth and Sports (Project: MSCA Fellowships CZ FZU II - CZ.02.01.01/00/22\_010/0008124).

\end{document}